\documentclass[conference]{IEEEtran}

\usepackage{amsmath,amssymb,amsfonts}
\usepackage{amsthm}
\usepackage{graphicx}
\usepackage{booktabs}
\usepackage{multirow}
\usepackage{algorithm}
\usepackage{algorithmic}
\usepackage{xcolor}
\usepackage{cite}
\usepackage{url}
\usepackage{xurl}
\usepackage{tikz}
\usetikzlibrary{positioning,fit,arrows.meta,calc}
\usepackage{hyperref}
\usepackage{placeins}
\usepackage{subcaption}
\usepackage{balance}

\newcommand{\figplaceholder}[1]{%
\fbox{\parbox[c][1.35in][c]{0.92\linewidth}{\centering #1}}%
}

\title{CFD-Guided Detection of Concept Drift in Multimodal Physiologic Signals}

\author{%
\IEEEauthorblockN{Farouk Ganiyu Adewumi\textsuperscript{\ensuremath{\parallel}}, Timothy Oladunni\textsuperscript{\P}, Rochak Ghimire\textsuperscript{*},\\
Kosisochukwu Ogbuanya\textsuperscript{\textdagger}, Sanaa Reeves\textsuperscript{\textdaggerdbl}, and Sandy Akoy\textsuperscript{\S}}
\IEEEauthorblockA{\textsuperscript{\ensuremath{\parallel}}Computer Science, Morgan State University, Baltimore, USA; fagan1@morgan.edu\\
\textsuperscript{\P}Computer Science, Morgan State University, Baltimore, USA; timothy.oladunni@morgan.edu\\
\textsuperscript{*}Computer Science, Morgan State University, Baltimore, USA; roghi2@morgan.edu\\
\textsuperscript{\textdagger}Computer Science, Fisk University, Nashville, USA; kosisoogbuanya@gmail.com\\
\textsuperscript{\textdaggerdbl}Electrical Engineering, Morgan State University, Baltimore, USA; sanareeves1@gmail.com\\
\textsuperscript{\S}Management Information Systems, University of Houston, Houston, USA; akoysandy@gmail.com}}

\begin{document}

\bstctlcite{IEEEexample:BSTcontrol}

\maketitle

\begin{abstract}
Cardiovascular AI models can classify clean electrocardiogram (ECG) signals, but real wearable signals change because of motion, breathing, posture, sensor contact, and true clinical deterioration. This paper asks when a model should keep its prediction, change it, or flag uncertainty. We propose a physiologic stability framework, called PECS, that compares changes inside the model with measurable changes in the signal. ECG is treated as the main cardiac signal, photoplethysmography (PPG) adds pulse and vascular information, and respiration is used only when ECG and PPG disagree. We test the framework on PTB-XL at pilot and full scales and on synchronized BIDMC and MIMIC waveform cohorts. The PTB-XL pilot and full-scale analyses selected different domain pairs, and the strongest cross-modal pair also changed across BIDMC and MIMIC, showing that adding every available signal is not always the best choice. PECS outperformed the evaluated drift-detection baseline implementations, reaching drift classification accuracy (DCA) of 0.8786 on expanded BIDMC and 0.9560 on MIMIC. The MIMIC results also showed that respiration can help during disagreement cases, but it should be used selectively rather than as an automatic override. Overall, the results support PECS as a candidate monitoring framework for wearable cardiovascular AI while highlighting the need for scale-aware domain selection and interpretable trust routing.
\end{abstract}

\begin{IEEEkeywords}
cardiovascular disease, electrocardiogram, photoplethysmography, concept drift, complementary feature domains, multimodal learning, physiologic signals, explainable AI
\end{IEEEkeywords}

\section{Introduction}
Cardiovascular disease (CVD) remains the leading cause of mortality worldwide, motivating the development of automated systems that can detect clinically meaningful changes from continuously measured physiologic signals \cite{who_cvd_2024,fuadah2025advances}. Deep learning has shown strong performance for ECG interpretation on large datasets such as PTB-XL \cite{PhysioNet-ptb-xl-1.0.3,strodthoff2021deep}, and related work has demonstrated the value of multimodal ECG representations, explainable fusion, and domain complementarity for biomedical signal classification \cite{oladunni2025rethinking,oladunni2025explainable}. However, high retrospective accuracy does not by itself solve the deployment problem.

In real environments, ECG and PPG signals are nonstationary. Motion, posture, respiration, medication, heart rate variation, vascular tone, and sensor contact can change the waveform without changing the underlying diagnosis. Conversely, early myocardial infarction, conduction changes, or arrhythmic transitions may produce signal changes that should alter the model's clinical interpretation. A deployed model therefore needs a principled way to decide whether a new signal segment represents benign drift, non-benign drift, or uncertainty.

This paper addresses that gap by presenting and evaluating a CFD-guided PECS framework. The framework separates three roles. ECG provides the primary cardiovascular prediction. PPG validates whether the ECG drift is physiologically consistent through a mechanically coupled but distinct measurement channel. Respiration is not used as a CVD predictor; it is used only as a tiebreaker when ECG and PPG disagree. The study is guided by the following problem statement:

\textit{Problem statement:} Existing ECG-based cardiovascular AI models can classify disease from curated signals, but they lack a reliable mechanism for deciding whether predictions should be held, updated, or flagged when deployed under physiologic drift. Multimodal fusion may help, but adding ECG, PPG, and respiratory information without domain selection can introduce redundancy, unstable routing, or misleading confidence.

We test three hypotheses:

\textit{H1:} CFD analysis can identify ECG-PPG domain pairs that provide complementary, non-redundant information for drift-aware cardiovascular prediction.

\textit{H2:} PECS energy-geometric stability improves benign versus non-benign drift classification over the evaluated generic drift baselines.

\textit{H3:} CFD-guided multimodal trust arbitration improves reliability and morphology preservation over non-CFD fusion, while respiratory tiebreaking improves disagreement handling when ECG and PPG conflict.

The contributions of this work are:
\begin{itemize}
    \item a problem formulation for deciding when a biomedical model should change its prediction under physiologic drift;
    \item a CFD-guided domain discovery workflow over ECG and PPG time, frequency, and time-frequency domains;
    \item a PECS stability rule connecting latent geometric drift to relative physiologic energy change;
    \item a scale comparison between a 2,100-record PTB-XL pilot and a 21,000-record full-scale rerun, followed by 5,035-window BIDMC and 5,000-segment MIMIC analyses showing how CFD rankings, drift classification, reliability gain, morphology preservation, and trust hierarchy behavior change with dataset source and analysis scale;
    \item a perturbation-stability analysis showing that top-ranked complementary pairs are more robust than low-ranked frequency-heavy pairs under realistic signal distortions.
\end{itemize}

The remainder of the paper follows the theory-to-evidence organization used in the CFD-GAN reference study: Section~II establishes the clinical and methodological background; Section~III defines the CFD-guided PECS framework; Section~IV specifies the datasets, baselines, and evaluation protocol; Section~V reports experiments in scale order; and Sections~VI--VIII present the discussion, limitations, and conclusion.

\section{Background and Related Work}
\subsection{ECG-Based Cardiovascular Prediction}
Large-scale ECG datasets and deep learning benchmarks have made automated cardiovascular interpretation more reproducible. PTB-XL provides a publicly available clinical 12-lead ECG dataset with diagnostic labels and standardized folds \cite{PhysioNet-ptb-xl-1.0.3}. Strodthoff et al. benchmarked deep learning methods on PTB-XL and showed that representation choice matters for ECG classification \cite{strodthoff2021deep}. MIMIC-derived waveform resources extend this setting toward intensive-care physiologic monitoring, where synchronized ECG, PPG, respiration, and blood-pressure signals make multimodal drift questions more realistic \cite{PhysioNet-mimic3wdb-matched-1.0,liang2018mimic,moulaeifard2026mimicppg}. Broader ECG deep learning studies have also demonstrated strong arrhythmia and diagnostic performance, but these models are typically evaluated under static train-test assumptions rather than under physiologic drift \cite{hannun2019cardiologist,ribeiro2020automatic}.

\subsection{Multimodal Physiologic Learning}
Multimodal cardiovascular AI is attractive because ECG, PPG, and respiration observe coupled aspects of cardiac function. ECG measures electrical activation, whereas PPG measures downstream vascular volume changes. PPG has become especially important for wearable cardiovascular monitoring because it is inexpensive, widespread, and compatible with smart devices \cite{charlton2022wearable}. Recent reviews highlight both the promise and the risks of combining ECG, PPG, phonocardiography, and other physiologic modalities for CVD classification \cite{fuadah2025advances,gawande2025survey,rathnayake2026wearable}. ECG-PPG translation and cross-modal representation studies further show that PPG can carry cardiac information, but they also reinforce the need to distinguish useful coupling from redundant or misleading fusion \cite{khuong2023ppgtoecg,fang2025ppgflowecg,liu2026biosignal}. However, multimodality is not automatically beneficial. If domains are redundant or weakly predictive, fusion can increase dimensionality without improving clinical information.

Oladunni and Wong formalized this concern through Complementary Feature Domain theory, arguing that useful multimodal biomedical fusion depends on complementarity rather than the number of domains \cite{oladunni2025rethinking}. Related explainable ECG work further compared intermediate and late fusion strategies and showed that fusion design affects both performance and interpretability \cite{oladunni2025explainable}. The present work extends this logic from ECG-domain fusion to ECG-PPG cross-modal trust validation.

\subsection{Concept Drift and Model Degradation}
Concept drift describes changes in the relationship between features and labels over time. Surveys by Gama et al. and Lu et al. provide the general machine learning foundation for drift detection and adaptation \cite{gama2014survey,lu2018learning}. In healthcare, drift is especially difficult because distribution shift can reflect either nuisance variation or genuine clinical change. Recent work on model degradation emphasizes that deployed models need performance-aware monitoring rather than one-time validation \cite{bayram2022concept}. PECS builds on this idea but adds a physiologic constraint: benign drift should produce latent change that is proportional to measurable energy change in the signal \cite{oladunni2026energy,oladunni2025physiologic}.

\subsection{PPG and Respiratory Validation}
PPG is useful for cardiovascular monitoring, but it is not an oracle. Peripheral vasoconstriction, motion, posture, perfusion changes, and sensor contact can alter PPG independently of ECG pathology \cite{charlton2022wearable,lu2009comparison}. Similarly, respiration affects both ECG and PPG morphology and can provide clinically meaningful context during disagreement. Respiratory rate estimation from ECG and PPG has been widely studied \cite{charlton2018breathing,charlton2016assessment,liu2019respiratory,koumpouzi2026respiration}, and respiratory quality control has been proposed for derived ECG/PPG respiratory data \cite{birrenkott_resp_quality}. This paper uses respiration differently: as a trust-gate signal rather than as a derived replacement for ECG or PPG.

\subsection{Morphology Preservation and Complementarity}
Clinical signal quality depends on morphology. In ECG, amplitude, waveform timing, frequency structure, and time-frequency transients can encode disease-relevant information. The CFD-GAN work argues that synthetic ECG generation must preserve time, frequency, and time-frequency morphology, and that complementarity constraints can reduce collapse across representations \cite{oladunni2026cfdgan}. The PECS framework uses a similar motivation for drift: a trustworthy model should not only classify labels, but should preserve stable morphology under benign perturbation and respond to genuine clinical state changes.

Table~\ref{tab:related_work_gap} positions this study against the nearest methodological areas. Prior work establishes strong ECG classification, wearable multimodal learning, general concept-drift detection, and respiratory estimation, but does not jointly address physiologic drift classification, complementary domain selection, and conditional cross-modal trust routing.

\begin{table*}[t]
\caption{Methodological scope of related cardiovascular AI research.}
\label{tab:related_work_gap}
\centering
\scriptsize
\setlength{\tabcolsep}{4pt}
\renewcommand{\arraystretch}{1.18}
\begin{tabular}{p{1.55in}p{1.35in}p{1.35in}p{2.25in}}
\toprule
\textbf{Research area} & \textbf{Typical signals} & \textbf{Primary objective} & \textbf{Remaining gap addressed here} \\
\midrule
ECG classification \cite{PhysioNet-ptb-xl-1.0.3,strodthoff2021deep,hannun2019cardiologist,ribeiro2020automatic} & ECG & Static diagnostic prediction & Does not determine whether a prediction change under deployment drift is physiologically warranted. \\
Wearable multimodal learning \cite{charlton2022wearable,fuadah2025advances,khuong2023ppgtoecg} & ECG, PPG, and related sensors & Fusion, translation, or risk prediction & Usually assumes that adding modalities is beneficial rather than testing domain complementarity. \\
Concept-drift monitoring \cite{gama2014survey,lu2018learning,bayram2022concept} & General feature streams & Detect distribution or performance change & Does not distinguish nuisance physiologic variation from clinically relevant waveform change. \\
Respiratory estimation \cite{charlton2018breathing,charlton2016assessment,liu2019respiratory} & ECG, PPG, and respiration & Estimate respiratory rate or quality & Does not evaluate respiration as a conditional trust signal during ECG--PPG disagreement. \\
This work & ECG, PPG, and respiration & CFD-guided PECS drift classification & Tests domain selection, energy--geometric consistency, and conditional trust routing together. \\
\bottomrule
\end{tabular}
\end{table*}

\section{CFD-Guided PECS Framework}
This section moves from domain selection to drift decisions and then to multimodal arbitration. The order mirrors the framework-first progression of the CFD-GAN reference paper: CFD identifies informative, non-redundant representations; PECS evaluates whether change is physiologically consistent; morphology metrics quantify representation stability; and the trust hierarchy resolves cross-modal disagreement.

\subsection{CFD Domain Discovery}
CFD theory defines complementarity through three conditions: low redundancy, individual predictive value, and positive joint information gain \cite{oladunni2025rethinking}. In this work, a domain pair \((X,Y)\) is considered useful when
\[
I(X;Y)\rightarrow 0,
\]
\[
I(X;C)>0,\qquad I(Y;C)>0,
\]
and
\[
I(X,Y;C)>\max(I(X;C),I(Y;C)).
\]
The empirical CFD score is written as
\[
CFD(X,Y)=I(X,Y;C)-I(X;Y),
\]
where high task information and low redundancy produce a larger score. pyCFD 3.0 operationalizes this with fusion gain, redundancy scoring, bootstrap confidence intervals, and permutation testing.

\subsection{PECS Energy-Geometric Stability}
Let \(x_m\) be an original segment from modality or encoder \(m\), and let \(\tilde{x}_m\) be a perturbed segment. The encoder outputs are
\[
z_m=E_m(x_m),\qquad \tilde{z}_m=E_m(\tilde{x}_m).
\]
Signal energy is
\[
E(x_m)=\frac{1}{N}\sum_{n=1}^{N}x_m[n]^2,
\]
relative energy change is
\[
\Delta E_m=\frac{|E(\tilde{x}_m)-E(x_m)|}{E(x_m)+\epsilon},
\]
and latent geometric drift is
\[
\delta_m=\|z_m-\tilde{z}_m\|_2.
\]
The PECS decision rule is
\[
\Phi_m(x_m,\tilde{x}_m)=
\begin{cases}
1, & \delta_m\leq \kappa_m\Delta E_m \text{ and } \delta_m<\gamma_m(z_m),\\
0, & \text{otherwise},
\end{cases}
\]
where \(\Phi_m=1\) denotes benign drift, \(\Phi_m=0\) denotes non-benign drift, \(\kappa_m\) is an encoder-specific energy-sensitivity constant, and \(\gamma_m(z_m)\) is the classifier margin.

Algorithm~\ref{alg:pecs} summarizes the complete PECS decision path. CFD is used before deployment to select the ECG--PPG domain pair; PECS then evaluates energy--geometric consistency for each incoming window and invokes respiration only when the two primary modalities disagree.

\begin{algorithm}[!t]
\caption{CFD-Guided PECS Drift Decision for One Window}
\label{alg:pecs}
\footnotesize
\begin{algorithmic}[1]
\REQUIRE Reference/current windows \(x_m,\tilde{x}_m\), \(m\in\{\mathrm{ECG},\mathrm{PPG}\}\); selected domains \(d_m\); encoders \(E_m\); \(\kappa_m\); margins \(\gamma_m\); respiration \(r\); \(\epsilon>0\)
\ENSURE Decision \(y^{drift}\in\{\mathrm{benign},\mathrm{nonbenign},\mathrm{uncertain}\}\) and action \(a\in\{\mathrm{hold},\mathrm{update},\mathrm{flag}\}\)
\FOR{each \(m\in\{\mathrm{ECG},\mathrm{PPG}\}\)}
    \STATE Project \(x_m,\tilde{x}_m\) into selected domain \(d_m\)
    \STATE \(z_m\leftarrow E_m(x_m)\), \(\tilde z_m\leftarrow E_m(\tilde{x}_m)\)
    \STATE \(\Delta E_m\leftarrow |E(\tilde{x}_m)-E(x_m)|/(E(x_m)+\epsilon)\)
    \STATE \(\delta_m\leftarrow\|z_m-\tilde z_m\|_2\)
    \STATE \(\Phi_m\leftarrow\mathbb{1}[\delta_m\leq\kappa_m\Delta E_m\land\delta_m<\gamma_m(z_m)]\)
\ENDFOR
\IF{\(\Phi_{\mathrm{ECG}}=\Phi_{\mathrm{PPG}}=1\)}
    \STATE \(y^{drift}\leftarrow\mathrm{benign}\); \(a\leftarrow\mathrm{hold}\)
\ELSIF{\(\Phi_{\mathrm{ECG}}=\Phi_{\mathrm{PPG}}=0\)}
    \STATE \(y^{drift}\leftarrow\mathrm{nonbenign}\); \(a\leftarrow\mathrm{update}\)
\ELSIF{respiratory quality is adequate}
    \STATE \(R\leftarrow\mathrm{RespiratoryGate}(r)\)
    \IF{\(R=1\)}
        \STATE \(y^{drift}\leftarrow\mathrm{benign}\); \(a\leftarrow\mathrm{hold}\)
    \ELSE
        \STATE \(y^{drift}\leftarrow\mathrm{nonbenign}\); \(a\leftarrow\mathrm{update}\)
    \ENDIF
\ELSE
    \STATE \(y^{drift}\leftarrow\mathrm{uncertain}\); \(a\leftarrow\mathrm{flag}\)
\ENDIF
\RETURN \(y^{drift},a\)
\end{algorithmic}
\end{algorithm}

\subsection{Morphology Preservation}
The latent representation is decomposed into a stable morphology component and a change-sensitive component:
\[
z_m=z^{stable}_m+z^{delta}_m.
\]
Morphology preservation is measured by
\[
MPI=\cos(z^{stable}_{before},z^{stable}_{after}).
\]
The morphology preservation gain is
\[
MPG=MPI_{CFD}-MPI_{nonCFD}.
\]

\subsection{Trust Hierarchy}
PECS computes independent ECG and PPG drift decisions, \(\Phi_{ECG}\) and \(\Phi_{PPG}\). If the two agree, the framework trusts the shared decision. If they disagree, respiration is used as a tiebreaker:
\[
R(t)=
\begin{cases}
1, & \text{normal respiratory pattern},\\
0, & \text{abnormal respiratory pattern}.
\end{cases}
\]
Here, abnormal includes tachypnea, bradypnea, irregular rhythm, or apnea.
The Tiebreaker Resolution Accuracy (TRA) is
\[
TRA=\frac{1}{|\mathcal{D}_0|}\sum_{i\in \mathcal{D}_0}
\mathbb{1}[R_i(t)=y_i^{drift}],
\]
where \(\mathcal{D}_0\) is the set of ECG-PPG disagreement cases.

\section{Experimental Configuration}
The evaluation is specified only after the complete framework has been defined. This separates the proposed mechanism from the datasets and operating points used to test it, matching the framework--configuration--results progression of the reference paper.

\subsection{Data Sources}
Three public waveform resources supplied the experimental evidence. PTB-XL contains 21,837 clinical 12-lead, 10-s ECG records; this study used a 2,100-record pilot and 21,000 usable records after full-scale preprocessing \cite{PhysioNet-ptb-xl-1.0.3,oladunni2026cfdgan}. BIDMC contains 53 eight-minute recordings with synchronized lead-II ECG, PPG, and impedance respiration sampled at 125~Hz \cite{7748483}. The MIMIC-III Waveform Database Matched Subset provides synchronized ICU waveform records linked to clinical records, including ECG, PPG, and respiration when simultaneously available \cite{PhysioNet-mimic3wdb-matched-1.0}. Table~\ref{tab:datasets} separates each published resource from the analysis cohort derived from it.

ECG and PPG were decomposed into time, frequency, and time-frequency representations, producing six candidate domains:
\[
\{ECG_t, ECG_f, ECG_{tf}, PPG_t, PPG_f, PPG_{tf}\}.
\]
The cross-modal pyCFD experiments evaluated all 15 pairwise combinations with 50 epochs, batch size 32, patience 8, 200 bootstrap samples, and 200 permutation tests.

\subsection{Dataset Roles and Analysis Scales}
For auditability, Table~\ref{tab:datasets} specifies the published dataset scope, the exact analytical scale, and the role of each cohort before any result is reported.

\begin{table*}[t]
\centering
\caption{Datasets, citations, analysis scales, and roles in the study. Published-resource counts are distinguished from the usable analysis cohorts.}
\label{tab:datasets}
\scriptsize
\renewcommand{\arraystretch}{1.18}
\begin{tabular}{p{0.17\linewidth}p{0.15\linewidth}p{0.25\linewidth}p{0.34\linewidth}}
\toprule
\textbf{Dataset} & \textbf{Signals} & \textbf{Published resource} & \textbf{Analysis cohort and role} \\
\midrule
PTB-XL \cite{PhysioNet-ptb-xl-1.0.3} & 12-lead ECG & 21,837 ten-second records from 18,885 patients & Small-scale pilot: 2,100 records. Full-scale rerun: 21,000 usable records. Both compare ECG time, frequency, and time-frequency domains; the full-scale rerun is the primary PTB-XL result. \\
BIDMC PPG and Respiration \cite{7748483} & Lead-II ECG, PPG, RESP & 53 eight-minute recordings sampled at 125~Hz & Feasibility analysis: 800 windows. Main analysis: 5,035 synchronized windows. Perturbation expansion: 35,000 cases for PECS, morphology, trust, and multimodal-advantage tests. \\
MIMIC-III Waveform Matched Subset \cite{PhysioNet-mimic3wdb-matched-1.0} & ECG, PPG, RESP & 22,317 waveform records matched to 10,282 clinical database records & Cross-dataset analysis: 5,000 synchronized segments, including a 1,250-segment held-out test partition. \\
\bottomrule
\end{tabular}
\end{table*}

Table~\ref{tab:datasets} clarifies three otherwise easy-to-confuse scales. The 2,100-record PTB-XL pilot is retained to show whether domain selection changes with sample size, whereas the 21,000-record rerun supports the main unimodal claim. BIDMC supplies tightly synchronized modalities for controlled perturbation and trust-routing tests, and MIMIC tests whether the cross-modal conclusions persist in a distinct ICU waveform source. Counts from the published resources are therefore not interchangeable with the smaller, quality-controlled cohorts used in the experiments.

\subsection{Comparative Analysis Protocol}
The paper treats the 2,100-record PTB-XL run as a small-scale pilot and the 21,000-record run as the full-scale unimodal ECG domain-discovery experiment. In the supplied pilot implementation, \texttt{max\_records=2100} limits the loader, and the generated report confirms 2,100 samples in each of the time, frequency, and time-frequency domain matrices. The pilot used same-architecture encoders with separately learned weights, 50 training epochs, a batch size of 32, early-stopping patience of 8, 200 bootstrap replicates, and 200 permutations. The 5,035-window BIDMC run is the primary cross-modal analysis, and the MIMIC run is a cross-dataset evaluation using the same ECG-PPG-RESP trust hierarchy. The comparison asks whether the same qualitative claims survive changes in dataset source, sample size, preprocessing, perturbation generation, and downstream PECS evaluation. The perturbation-stability analysis compares the top five and bottom five domain pairs under Gaussian noise, amplitude scaling, baseline drift, time masking, and combined perturbations. A finding is considered robust when (i) the pair remains positive under perturbation, (ii) its confidence interval remains largely above zero, and (iii) its rank remains stronger than fragile low-ranked pairs.

\subsection{Baseline Configurations}
For the reference drift comparisons, ADWIN \cite{bifet2007adwin} and Page-Hinkley \cite{xiang2023concept} used the default configurations in the River library and were applied to the same energy-change stream. MMD \cite{gretton2012kernel} used an RBF kernel with the median-distance bandwidth heuristic and 200 permutations, while Monte Carlo dropout uncertainty \cite{gal2016dropout} used 30 stochastic forward passes. These generic baselines were not tuned with the same task-specific physiologic structure as PECS; consequently, their results provide reference operating points for the evaluated implementations rather than evidence that PECS is intrinsically superior to every possible tuned version of these detectors.

\subsection{Perturbation Labels and Evaluation Scope}
The benign/non-benign targets in the perturbation experiments were assigned a priori from the perturbation family, not from independent clinical adjudication. Benign families represented intended nuisance variation, including amplitude scaling, baseline wander, respiratory modulation, Gaussian noise, depth or rate variation, and minor irregularity. Non-benign families represented simulated morphology or signal-failure events, including ST elevation, QRS widening, T-wave inversion, simulated atrial fibrillation, PPG amplitude or perfusion reduction, notch removal, pulselessness, and abnormal respiratory patterns. The label was fixed before computing relative energy change \(\Delta E_m\), latent drift \(\delta_m\), or the PECS decision; therefore, the target was not numerically generated by the PECS rule. Nevertheless, the perturbation taxonomy and PECS share physiologic assumptions about which changes should be consequential. DCA should consequently be interpreted as agreement with a controlled synthetic perturbation taxonomy, not as accuracy against independently adjudicated clinical deterioration.

\section{Results}
Table~\ref{tab:experiment_summary} provides a reading map from the seven result blocks to their scientific questions and primary conclusions. The detailed subsections then report the evidence in the same order, beginning with scale sensitivity in PTB-XL and ending with cross-dataset evaluation in MIMIC.

\begin{table*}[t]
\centering
\caption{Summary of experiments, questions, and primary outcomes.}
\label{tab:experiment_summary}
\small
\renewcommand{\arraystretch}{1.18}
\begin{tabular}{p{0.07\linewidth}p{0.31\linewidth}p{0.24\linewidth}p{0.28\linewidth}}
\toprule
Exp. & Plain-language question & Dataset or test & Takeaway \\
\midrule
1 & Does the selected ECG pair change with scale? & PTB-XL pilot and full-scale cohorts & Time+Frequency in the pilot; Frequency+TimeFrequency at full scale \\
2 & Do selected ECG and PPG domains add useful information together? & BIDMC & Selected pairs outperformed full fusion \\
3 & Are strong pairs stable under noisy signals? & BIDMC and MIMIC perturbation tests & Strong pairs were more stable than weak pairs \\
4 & Does PECS beat the evaluated drift detectors? & BIDMC drift classification & PECS reached DCA = 0.8786 \\
5 & Does CFD improve reliability and morphology? & CFD vs non-CFD fusion & Reliability improved; morphology showed a trade-off \\
6 & Does respiration help when ECG and PPG disagree? & BIDMC trust analysis & Respiration helped selectively \\
7 & Do the findings transfer to MIMIC? & MIMIC cross-dataset evaluation & PECS reached DCA = 0.9560 \\
\bottomrule
\end{tabular}
\end{table*}

The sequence in Table~\ref{tab:experiment_summary} is intentional: PTB-XL isolates sample-scale effects within ECG, BIDMC adds synchronized PPG and respiration, and MIMIC tests transport to a different clinical source. Accordingly, results from different rows should be compared qualitatively rather than treated as a single pooled benchmark.

\subsection{Experiment 1: Small- and Full-Scale PTB-XL ECG Domain Discovery}
The 2,100-record pilot ranked Time+Frequency first (CFD Index = 3.5128, fusion gain = 0.0495, \(p=0.004975\), 95\% CI [0.0266, 0.0820]), followed by Time+TimeFrequency (CFD Index = 1.8269, fusion gain = 0.0210, \(p=0.024876\), 95\% CI [0.0057, 0.0362]). Both were classified as weakly complementary. Frequency+TimeFrequency was non-complementary (CFD Index = -0.3145, fusion gain = -0.0038, \(p=0.6119\), 95\% CI [-0.0286, 0.0267]). The pilot's three-domain fusion was also weakly complementary (CFD Index = 2.9033, fusion gain = 0.0362, \(p=0.034826\), 95\% CI [0.0076, 0.0724]), but Time+Frequency remained the recommended strategy. The full-scale rerun used 21,000 usable ECG records and reversed the pairwise ordering: Frequency+TimeFrequency became the only pair satisfying the prespecified gain, permutation, and confidence-interval criteria, with CFD Index = 0.7769, fusion gain = 0.0170, \(p=0.004975\), and 95\% CI [0.0069, 0.0255]. Table~\ref{tab:ptbxl_cfd} makes the two scales explicit and prevents the exploratory pilot result from being mistaken for the primary full-scale result.

\begin{table*}[!t]
\centering
\caption{Experiment 1: PTB-XL small- and full-scale ECG domain discovery. The selected pair at each scale is reported with its inferential statistics.}
\label{tab:ptbxl_cfd}
\small
\renewcommand{\arraystretch}{1.12}
\begin{tabular}{llp{0.22\textwidth}ccccp{0.13\textwidth}}
\toprule
\textbf{Scale} & \textbf{N} & \textbf{Selected pair} & \textbf{CFD} & \textbf{Gain} & \textbf{p} & \textbf{95\% CI} & \textbf{Decision} \\
\midrule
Small-scale pilot & 2,100 & Time+Frequency & 3.5128 & 0.0495 & 0.0050 & [0.0266, 0.0820] & Complementary \\
Full-scale rerun & 21,000 & Frequency+TimeFrequency & 0.7769 & 0.0170 & 0.0050 & [0.0069, 0.0255] & Complementary \\
\bottomrule
\end{tabular}
\end{table*}

Table~\ref{tab:ptbxl_cfd} shows that both selected pairs satisfy the statistical decision rule, yet their identities differ. Thus, the scale comparison concerns which ECG representations complement one another, not whether the corrected pilot produced a significant result. The full-scale evidence supports Frequency+TimeFrequency for the primary analysis, while the pilot is retained as a scale-sensitivity result.

\begin{figure*}[!t]
\centering
\begin{subfigure}[t]{0.47\textwidth}
\centering
\IfFileExists{figures/ptbxl_cfd_matrix.png}{\includegraphics[width=\linewidth]{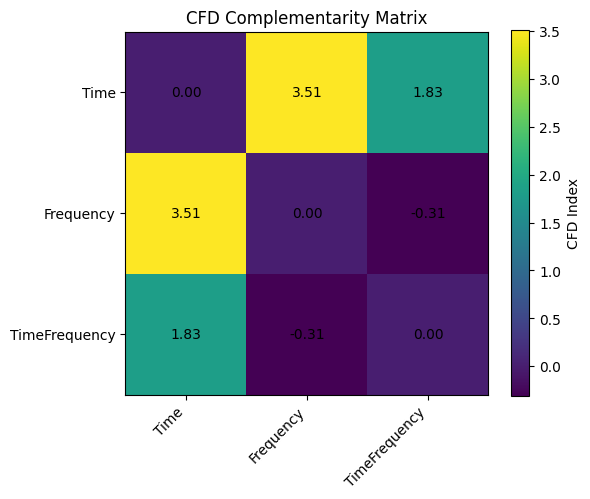}}{\figplaceholder{Upload PTB-XL small-scale CFD matrix here as figures/ptbxl_cfd_matrix.png}}
\caption{Small-scale pilot (2,100 records).}
\label{fig:ptbxl_small_cfd}
\end{subfigure}\hfill
\begin{subfigure}[t]{0.47\textwidth}
\centering
\IfFileExists{figures/ptbxl_full_cfd_matrix.png}{\includegraphics[width=\linewidth,trim=0.8in 0.25in 0.8in 0.2in,clip]{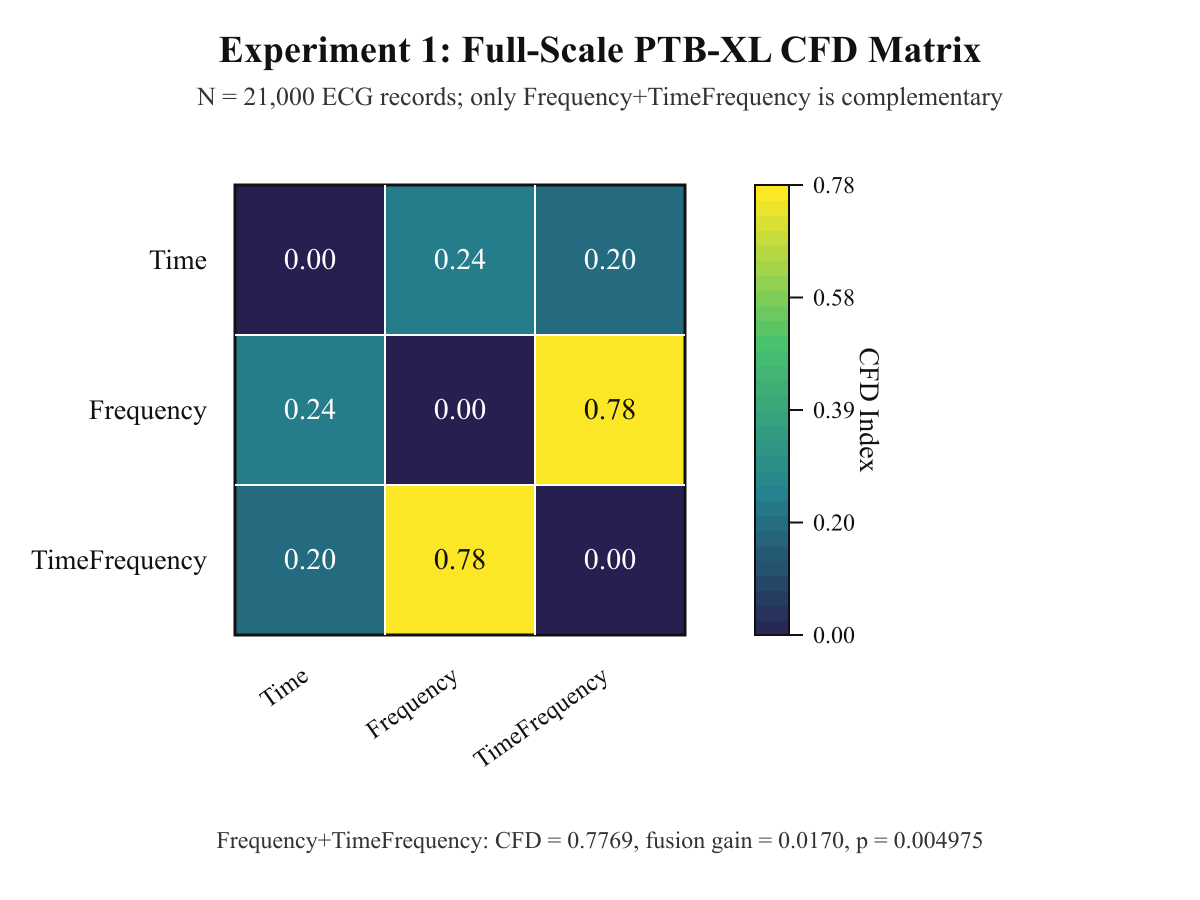}}{\figplaceholder{Upload PTB-XL full-scale CFD matrix here as figures/ptbxl_full_cfd_matrix.png}}
\caption{Full-scale rerun (21,000 usable records).}
\label{fig:ptbxl_full_cfd}
\end{subfigure}
\caption{PTB-XL CFD complementarity matrices at two analysis scales. The dominant pair changes from Time+Frequency in the pilot to Frequency+TimeFrequency in the full-scale rerun.}
\label{fig:ptbxl_scale}
\end{figure*}

\begin{figure}[!t]
\centering
\IfFileExists{figures/ptbxl_full_cfd_ranking.png}{\includegraphics[width=\linewidth]{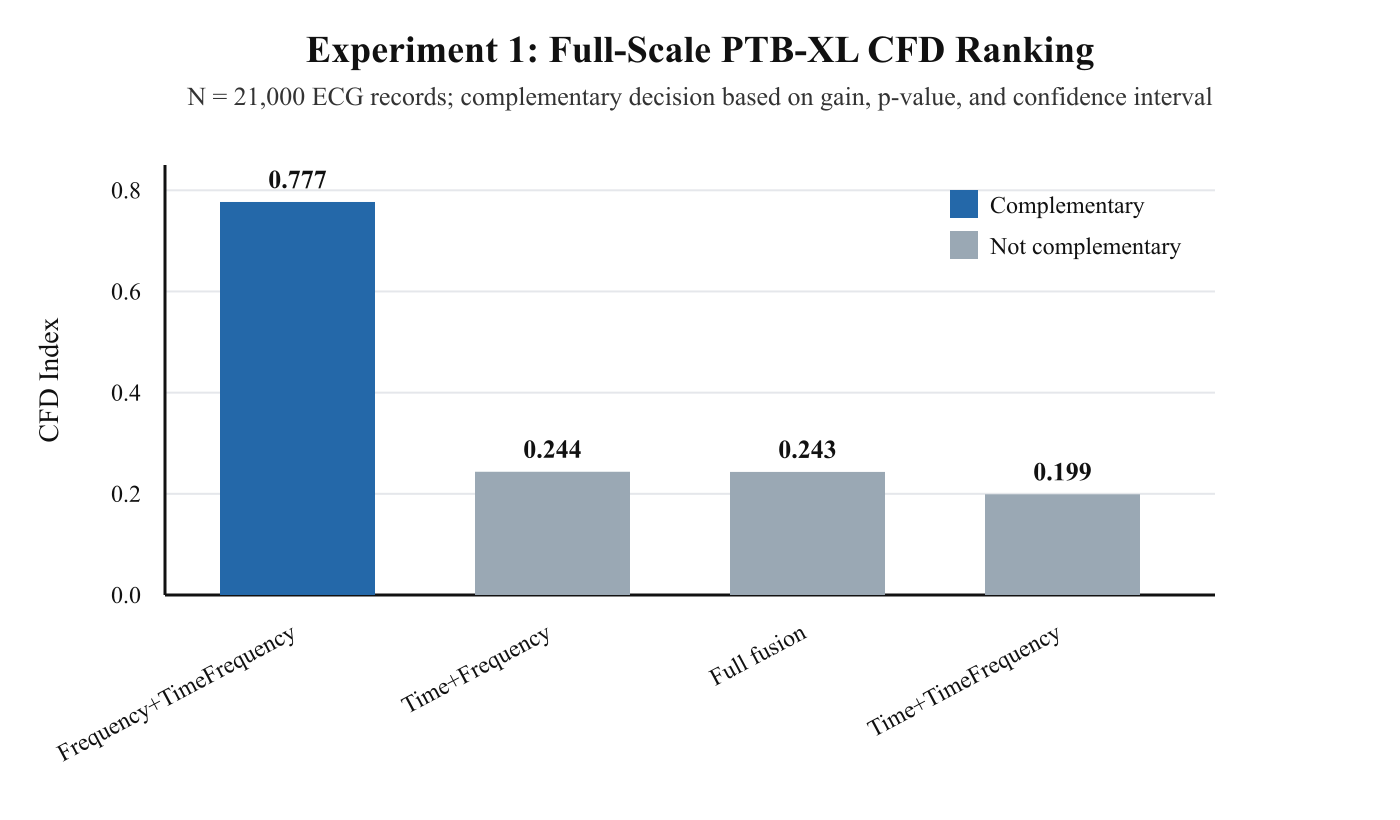}}{\figplaceholder{Upload PTB-XL full-scale CFD ranking here as figures/ptbxl_full_cfd_ranking.png}}
\caption{Experiment 1: full-scale PTB-XL CFD combination ranking.}
\label{fig:ptbxl_full_ranking}
\end{figure}

Figure~\ref{fig:ptbxl_scale} visually confirms the rank reversal across scales. Specifically, Fig.~\ref{fig:ptbxl_small_cfd} concentrates its largest pilot score in Time+Frequency, whereas Fig.~\ref{fig:ptbxl_full_cfd} concentrates its largest full-scale score in Frequency+TimeFrequency. Figure~\ref{fig:ptbxl_full_ranking} adds the inferential decision layer: only Frequency+TimeFrequency is marked complementary, while the other pairs and full fusion fail at least one prespecified criterion. Together, the figures justify using the full-scale pair downstream and caution against freezing an architecture from a small pilot.

\subsection{Experiment 2: Cross-Modal ECG-PPG CFD Discovery}
The 5,035-window BIDMC analysis showed that selected cross-modal ECG-PPG pairs were more useful than indiscriminate full fusion. The recommended pair was ECG-Frequency+PPG-Time-Frequency (CFD Index = 0.5261, fusion gain = 0.0183, \(p=0.004975\)), followed by ECG-Time+PPG-Time (CFD Index = 0.3237, fusion gain = 0.0191, \(p=0.009950\)) and ECG-Frequency+PPG-Time (CFD Index = 0.3101, fusion gain = 0.0143, \(p=0.029851\)). Full six-domain fusion remained non-complementary (CFD Index = 0.1074, \(p=0.218905\)). Table~\ref{tab:expanded_cfd} shows that the decision is not determined by CFD magnitude alone: positive gain and the permutation criterion must also be satisfied. The earlier 800-window feasibility analysis and its complete ranking are reported in the supplementary material.

\begin{table}[!t]
\centering
\caption{Experiment 2: BIDMC cross-modal CFD comparison.}
\label{tab:expanded_cfd}
\scriptsize
\setlength{\tabcolsep}{3pt}
\begin{tabular}{lcccc}
\toprule
Pair & CFD & Gain & \(p\) & Decision \\
\midrule
ECG-Freq+PPG-TF & 0.5261 & 0.0183 & 0.0050 & Yes \\
ECG-Time+PPG-Time & 0.3237 & 0.0191 & 0.0100 & Yes \\
ECG-Freq+PPG-Time & 0.3101 & 0.0143 & 0.0299 & Yes \\
ECG-Time+PPG-Freq & 0.2708 & 0.0103 & 0.0746 & No \\
Full fusion & 0.1074 & 0.0048 & 0.2189 & No \\
ECG-TF+PPG-TF & -0.1254 & -0.0048 & 0.8358 & No \\
\bottomrule
\end{tabular}
\end{table}

The contrast in Table~\ref{tab:expanded_cfd} is important for architecture selection. Three targeted cross-modal pairs met the decision rule, but the six-domain model and ECG-TF+PPG-TF did not; therefore, more inputs did not translate into stronger complementary information.

\subsection{Experiment 3: Perturbation Stability of Complementary Pairs}
The perturbation analysis evaluated whether complementarity survives Gaussian noise, amplitude scaling, baseline drift, time masking, and combined distortions. Top-ranked pairs retained higher mean CFD values, narrower confidence intervals, and positive sign stability more often than bottom-ranked pairs. Weak frequency-heavy combinations were especially fragile. This supports the hypothesis that strong complementary relationships are more perturbation-stable than weak or redundant relationships. Detailed exploratory rankings and heatmaps are provided in the supplementary material.

\subsection{Experiment 4: PECS Drift Classification}
In the BIDMC perturbation experiment, PECS achieved DCA = 0.8786, F1-score = 0.8783, EDC = 0.8446, ECG SVR = 2.63\%, and PPG SVR = 4.74\%. Here, DCA measures agreement with the prespecified perturbation-family taxonomy rather than independently adjudicated clinical deterioration. The analysis produced 14,348 ECG-PPG disagreement cases out of 35,000 perturbation cases. As summarized in Table~\ref{tab:dca_baselines}, PECS performed best against the evaluated baseline implementations, exceeding ADWIN by 44.88 percentage points, Page-Hinkley by 36.19 points, MMD by 8.55 points, and dropout uncertainty by 1.12 points. This supports H2 within the controlled perturbation setting and suggests that the PECS energy-geometric condition provides a stronger drift signal than the evaluated generic statistical detectors under this label definition.

\begin{table}[!t]
\centering
\caption{Experiment 4: drift classification accuracy against cited baseline methods.}
\label{tab:dca_baselines}
\begin{tabular}{lc}
\toprule
Method & DCA \\
\midrule
PECS & 0.8786 \\
ADWIN \cite{bifet2007adwin} & 0.4298 \\
Page-Hinkley \cite{xiang2023concept} & 0.5167 \\
MMD \cite{gretton2012kernel} & 0.7931 \\
Dropout uncertainty \cite{gal2016dropout} & 0.8674 \\
\bottomrule
\end{tabular}
\end{table}

Table~\ref{tab:dca_baselines} also shows that dropout uncertainty was a much stronger comparator than ADWIN or Page-Hinkley. The 1.12-point PECS advantage over dropout should therefore be interpreted more cautiously than the larger gaps against the stream-change detectors.

\subsection{Experiment 5: CFD Reliability and Morphology Preservation}
The CFD-optimal pair achieved DCA = 0.8786, the non-CFD pair achieved DCA = 0.8248, and the all-six-domain aggregation achieved DCA = 0.8087. The resulting CRG was 0.0538, with bootstrap \(p<0.0001\). Table~\ref{tab:crg_mpg} therefore indicates that CFD-guided selection improved drift reliability over non-CFD selection and over indiscriminate all-domain averaging.

Morphology preservation was less straightforward. The non-CFD ECG-TF encoder produced higher MPI under label-changing perturbations: ST elevation MPI was 0.9976 for CFD-optimal versus 0.9996 for non-CFD, and QRS widening MPI was 0.9597 for CFD-optimal versus 0.9888 for non-CFD. MPG was therefore negative for ST elevation (-0.0020), QRS widening (-0.0291), and on average (-0.0156). This difference suggests that high MPI can reflect insensitivity to clinically meaningful change rather than better preservation, reframing morphology preservation as a sensitivity--stability trade-off rather than a simple win for CFD.

\begin{table}[!t]
\centering
\caption{Experiment 5: CFD reliability and morphology preservation.}
\label{tab:crg_mpg}
\begin{tabular}{lcc}
\toprule
Architecture & DCA & MPI \\
\midrule
CFD-optimal & 0.8786 & 0.9597--0.9976 \\
Non-CFD & 0.8248 & 0.9888--0.9996 \\
All domains & 0.8087 & -- \\
\midrule
Gain & CRG = 0.0538 & MPG avg. = -0.0156 \\
\bottomrule
\end{tabular}
\end{table}

\begin{figure}[!t]
\centering
\IfFileExists{figures/morphology_preservation_mpi.png}{\includegraphics[width=\linewidth]{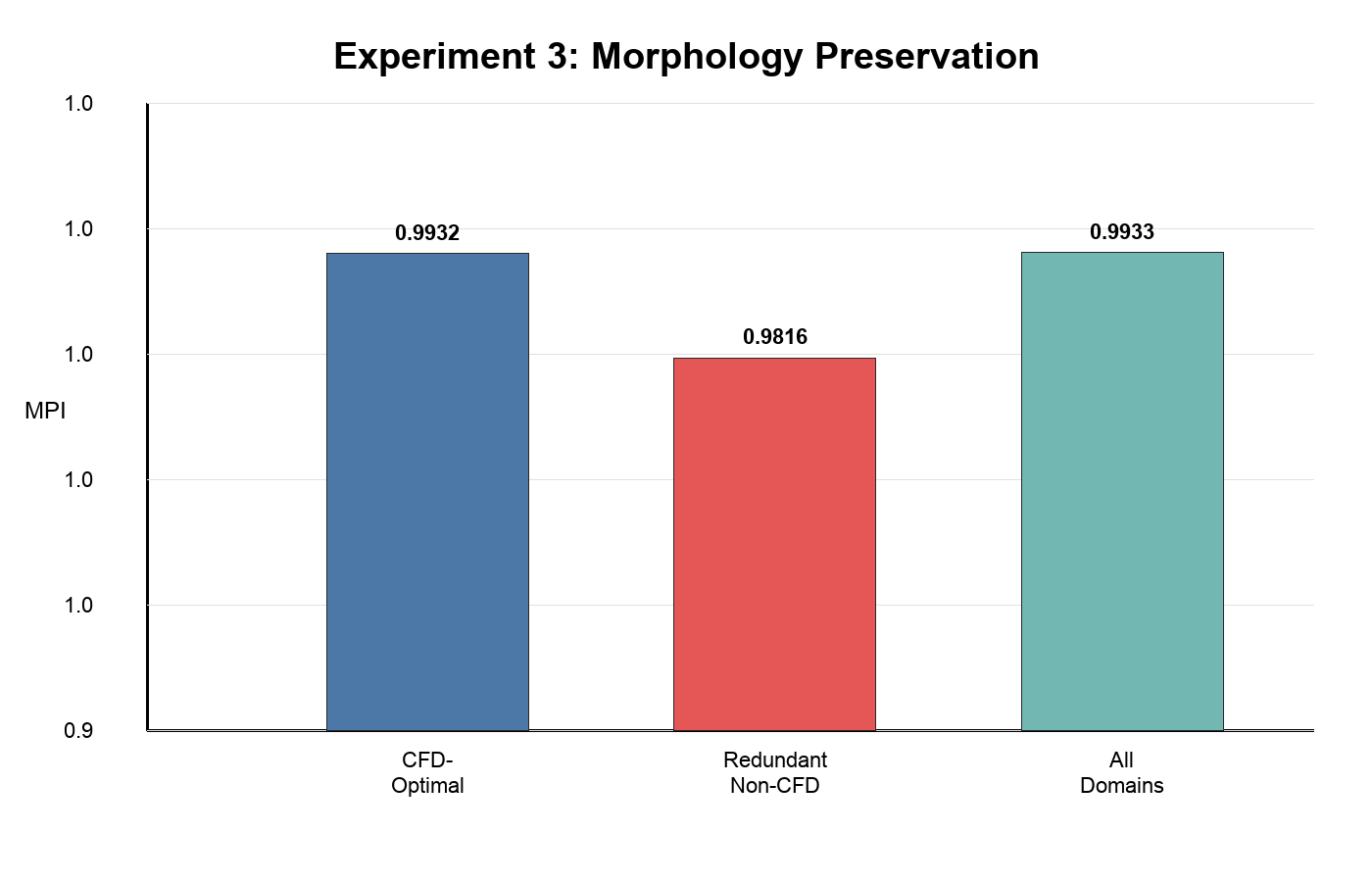}}{\figplaceholder{Upload MPI morphology preservation chart here as figures/morphology_preservation_mpi.png}}
\caption{Experiment 5: morphology preservation index comparison across CFD-optimal and non-CFD architectures.}
\label{fig:mpi}
\end{figure}

Figure~\ref{fig:mpi} makes the trade-off in Table~\ref{tab:crg_mpg} visible: the non-CFD encoder retains higher cosine similarity under ST elevation and QRS widening, even though it has lower drift-classification accuracy. The figure therefore does not support a blanket morphology-preservation advantage for CFD; instead, it suggests that the CFD-optimal encoder is more responsive to label-changing perturbations.

\subsection{Experiment 6: Trust Hierarchy and Respiratory Tiebreaking}
Out of 35,000 BIDMC cases, ECG and PPG agreed on 20,652 and disagreed on 14,348, giving CMAR = 0.5901. As reported in Table~\ref{tab:trust}, respiratory tiebreaking achieved TRA = 0.5133, below the 0.70 target, but improved disagreement-case DCA from 0.3511 to 0.5133. Thus, respiration provided useful information in disagreement cases but was not sufficiently accurate to serve as a standalone arbiter.

\begin{table}[!t]
\centering
\caption{Experiment 6: trust hierarchy metrics.}
\label{tab:trust}
\begin{tabular}{lc}
\toprule
Metric & Value \\
\midrule
CMAR & 0.5901 \\
TRA & 0.5133 \\
DCA without tiebreaker & 0.3511 \\
DCA with tiebreaker & 0.5133 \\
DCA gain & 0.1623 \\
Level 3 rate & 0.0000 \\
\bottomrule
\end{tabular}
\end{table}

The PPG failure analysis identified 1,814 false-benign PPG failures, 3,223 false-non-benign PPG failures, and 9,311 other disagreement cases. Respiration resolved 95.4\% of false-benign PPG failures but only 6.0\% of false-non-benign PPG failures; ECG fallback accuracy was 100\% for both failure groups in this experiment. This supports treating PPG as a secondary consistency signal rather than an oracle and suggests that respiratory arbitration is scenario-dependent rather than uniformly reliable.

\begin{figure}[!t]
\centering
\IfFileExists{figures/trust_hierarchy_flow.png}{\includegraphics[width=\linewidth]{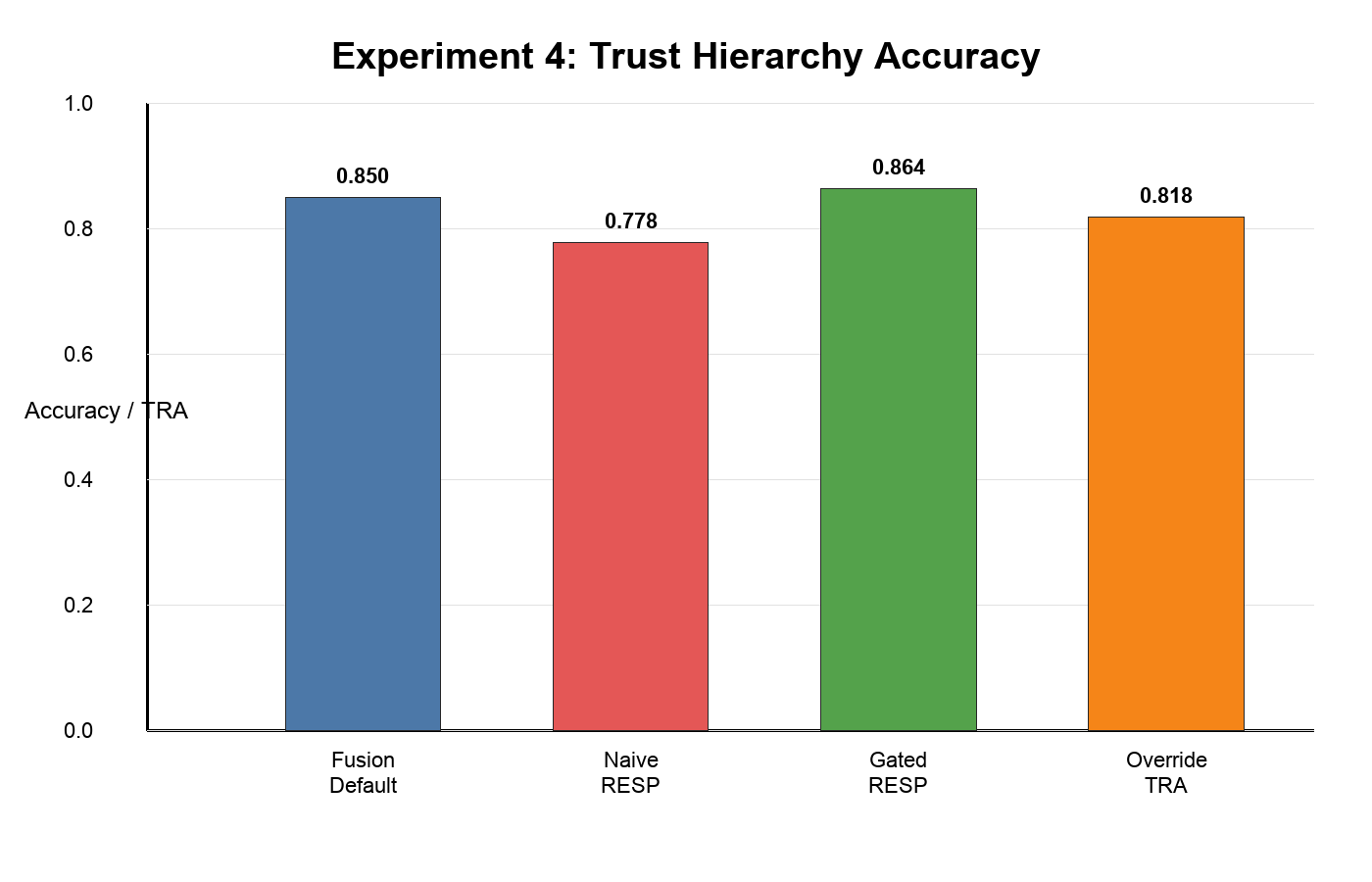}}{\figplaceholder{Upload trust hierarchy flow chart here as figures/trust_hierarchy_flow.png}}
\caption{Experiment 6: trust hierarchy flow showing agreement, disagreement, respiratory-resolved, and Level 3 cases.}
\label{fig:trust_flow}
\end{figure}

Figure~\ref{fig:trust_flow} locates that gain in the routing process: the large disagreement branch is where respiration can add information, while unresolved cases should remain at Level~3 rather than being forced into a confident decision. The zero Level~3 rate in Table~\ref{tab:trust} is therefore a limitation of this fixed rule, not evidence that uncertainty disappeared.

\subsection{Experiment 6B: Multimodal Drift Advantage}
When multimodal drift was computed from normalized ECG and PPG energy changes, unimodal ECG DCA was 0.7417, unimodal PPG DCA was 0.7885, and multimodal ECG+PPG DCA was 0.8741. Table~\ref{tab:mda} shows that this produced MDA = 0.1324 versus ECG and MDA = 0.0856 versus PPG, with bootstrap \(p<0.000001\). By trust level, MDA was slightly negative for Level~1 agreement cases (-0.0089) but strongly positive for Level~2 disagreement cases (0.2001). These results support a conditional view of multimodality: ECG+PPG fusion is most valuable when the modalities disagree and least necessary when they already agree.

\begin{table}[!t]
\centering
\caption{Experiment 6B: summary of multimodal drift advantage and OOD robustness.}
\label{tab:mda}
\begin{tabular}{lc}
\toprule
Metric & Value \\
\midrule
DCA ECG only & 0.7417 \\
DCA PPG only & 0.7885 \\
DCA multimodal & 0.8741 \\
MDA vs ECG & 0.1324 \\
MDA vs PPG & 0.0856 \\
MDA Level 2 & 0.2001 \\
\bottomrule
\end{tabular}
\end{table}

The Level~2 value in Table~\ref{tab:mda} is larger than either overall MDA comparison, confirming that the aggregate benefit is concentrated in disagreement cases rather than distributed uniformly across all samples.

\subsection{Experiment 7: MIMIC ECG-PPG-RESP Cross-Dataset Evaluation}
The MIMIC analysis added a third cross-modal physiologic dataset to the PTB-XL and BIDMC evidence. This run used 5,000 synchronized ECG, PPG, and respiratory segments and repeated the six-domain ECG-PPG CFD workflow. Unlike BIDMC, the MIMIC ranking returned ECG-Time+PPG-Time as the strongest pair, with CFD Index = 13.6544, fusion gain = 0.1416, \(p=0.004975\), and CI [0.1144, 0.1712]. The next strongest pairs were ECG-TF+PPG-Time (CFD Index = 12.5377), PPG-Time+PPG-Frequency (CFD Index = 10.5576), ECG-Frequency+PPG-Time (CFD Index = 8.9669), and ECG-Time+PPG-Frequency (CFD Index = 7.2028). The selected pair also matched the locked smaller-scale MIMIC selection, supporting the cross-modal complementarity claim while still showing that the best pair is dataset-dependent. Table~\ref{tab:mimic_summary} consolidates the ranking, drift, reliability, and trust results so that the selection claim is not separated from its downstream consequences.

\begin{table}[!t]
\centering
\caption{Experiment 7: MIMIC ECG-PPG-RESP evaluation summary.}
\label{tab:mimic_summary}
\small
\begin{tabular}{p{0.55\linewidth}p{0.35\linewidth}}
\toprule
Metric & Value \\
\midrule
Best CFD pair & ECG-Time+PPG-Time \\
CFD Index / gain & 13.6544 / 0.1416 \\
Fusion DCA & 0.9560 \\
ECG-only / PPG-only DCA & 0.9704 / 0.9360 \\
Fusion F1 / ROC-AUC & 0.9707 / 0.9906 \\
ADWIN / Page-Hinkley DCA & 0.5296 / 0.4972 \\
MMD / uncertainty DCA & 0.4980 / 0.4784 \\
CFD-optimal matched DCA & 0.9272 \\
Redundant-pair / all-domain DCA & 0.8072 / 0.9272 \\
Selected PECS CRG vs redundant & 0.1488 \\
Matched-architecture CRG vs redundant & 0.1200 \\
TRA on \(C(t)=0\) cases & 0.7588 \\
Manual trust hierarchy accuracy & 0.8704 \\
Overall MDA vs ECG & 0.2232 \\
\bottomrule
\end{tabular}
\end{table}

Table~\ref{tab:mimic_summary} reveals two qualifications. First, ECG-only DCA (0.9704) exceeded fusion DCA (0.9560), so the MIMIC result does not establish an unconditional fusion advantage. Second, the selected pair still improved substantially over the redundant pair and matched the all-domain model under the controlled architecture comparison, supporting selective fusion as a reliability--complexity trade-off.

On the 1,250-sample MIMIC test partition, the selected ECG-Time+PPG-Time PECS model achieved fusion DCA = 0.9560, balanced accuracy = 0.9409, F1 = 0.9707, and ROC-AUC = 0.9906. ECG alone reached DCA = 0.9704 and PPG alone reached DCA = 0.9360, but the fusion model gave the strongest cross-modal reliability profile and outperformed all required drift baselines: ADWIN-style detection (DCA = 0.5296), Page-Hinkley-style detection (DCA = 0.4972), MMD-style energy detection (DCA = 0.4980), and probability-uncertainty fallback (DCA = 0.4784). These gains ranged from 42.64 to 47.76 percentage points, with Bonferroni-corrected paired tests significant for all baseline comparisons.

For CFD reliability, the selected final PECS model achieved DCA = 0.9560, compared with 0.8072 for the redundant ECG-Frequency+PPG-Frequency baseline and 0.9272 for the all-domain model. This produced reporting CRG = 0.1488 versus the redundant baseline and CRG = 0.0288 versus all domains. The matched-architecture significance test isolated the domain-selection effect: CFD-optimal ECG-Time+PPG-Time reached DCA = 0.9272 versus 0.8072 for the redundant pair, CRG = 0.1200, bootstrap CI [0.0936, 0.1464], and McNemar \(p<10^{-6}\). Against all-domain fusion, matched-architecture DCA was identical at 0.9272, with CI [-0.0160, 0.0168] and McNemar \(p=0.9240\). Thus, MIMIC supports CFD-guided selection over redundant fusion while framing all-domain fusion as a complexity-performance trade-off.

The MIMIC CFD stability analysis added a useful robustness check. All five top-ranked pairs remained sign-stable, with confidence intervals fully above zero. The bottom five pairs did not become confidently complementary; their confidence intervals crossed zero, meaning their signs were unstable under resampling. The important distinction was therefore not lower measurement noise for strong pairs, but larger effect size relative to a similar bootstrap noise floor of roughly 0.008--0.015. Table \ref{tab:mimic_stability_assignment} summarizes this top-versus-bottom split.

\begin{table}[!t]
\centering
\caption{MIMIC n=5,000 CFD stability under bootstrap resampling.}
\label{tab:mimic_stability_assignment}
\scriptsize
\renewcommand{\arraystretch}{1.12}
\begin{tabular}{p{0.35\linewidth}ccc}
\toprule
\textbf{Pair} & \textbf{CFD} & \textbf{Gain CI} & \textbf{Stable?} \\
\midrule
ECG-Time+PPG-Time & 13.65 & [0.114, 0.171] & Yes \\
ECG-TF+PPG-Time & 12.54 & [0.104, 0.158] & Yes \\
PPG-Time+PPG-Freq & 10.56 & [0.082, 0.138] & Yes \\
ECG-Freq+PPG-Time & 8.97 & [0.066, 0.119] & Yes \\
ECG-Time+PPG-Freq & 7.20 & [0.048, 0.106] & Yes \\
\midrule
ECG-Freq+ECG-TF & -1.31 & [-0.038, 0.007] & No \\
ECG-Freq+PPG-TF & -0.32 & [-0.022, 0.015] & No \\
PPG-Time+PPG-TF & 0.43 & [-0.012, 0.024] & No \\
PPG-Freq+PPG-TF & 0.61 & [-0.007, 0.022] & No \\
ECG-Time+ECG-TF & 0.78 & [-0.010, 0.034] & No \\
\bottomrule
\end{tabular}
\end{table}

The stability analysis also clarifies why domain selection should be treated as relational rather than intrinsic. ECG-Time is the strongest domain when paired with PPG-Time, but it is not complementary when paired with ECG-TF. Similarly, ECG-Frequency is part of the weakest pair with ECG-TF but becomes the fourth-best pair when paired with PPG-Time. Six non-transitive triples were identified, usually routing through ECG-Time or PPG-Time as a bridging domain. In practical terms, a domain is not universally useful by itself; its value depends on the partner domain and dataset context.

\begin{figure}[!t]
\centering
\IfFileExists{figures/mimic_v2_pair_dca_ecg_presence.png}{\includegraphics[width=\linewidth]{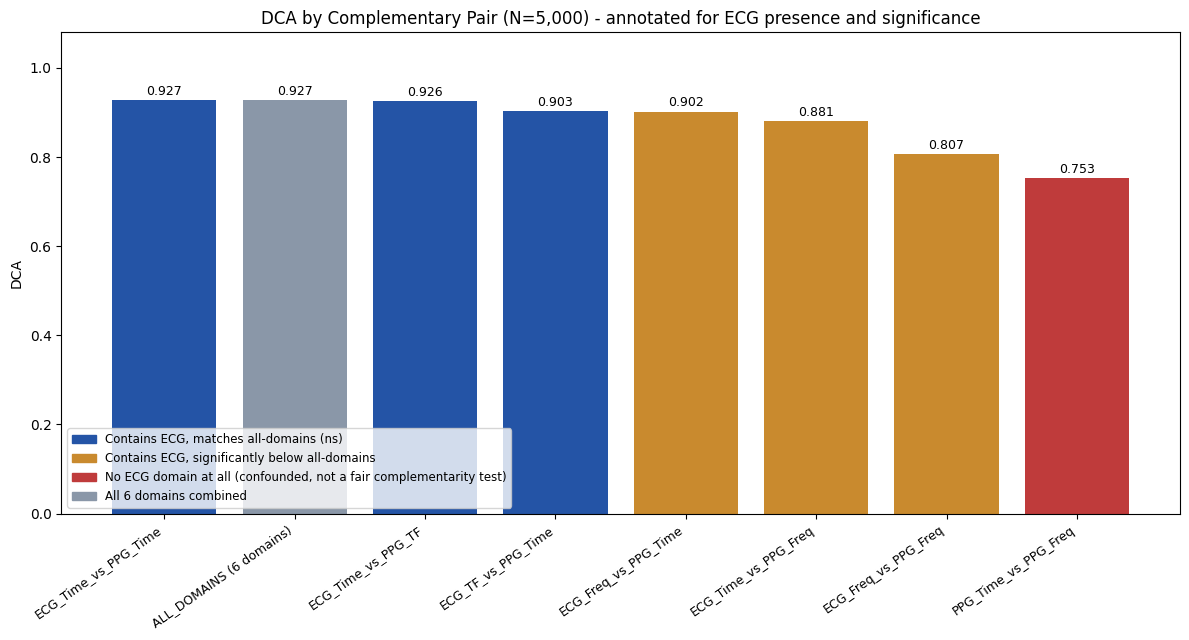}}{\figplaceholder{Upload MIMIC pairwise DCA chart here as figures/mimic_v2_pair_dca_ecg_presence.png}}
\caption{MIMIC n=5,000 DCA by complementary pair, annotated by ECG presence and significance. The top two-domain pair matches the all-domain model while using fewer inputs.}
\label{fig:mimic_v2_pair_dca_ecg_presence}
\end{figure}

Figure~\ref{fig:mimic_v2_pair_dca_ecg_presence} places the MIMIC pairwise DCA results beside the all-domain reference. The leading two-domain model reaches the same matched-architecture DCA as all-domain fusion with fewer inputs, whereas pairs without a useful ECG partner are generally weaker; this supports parsimony but not a universal pair choice.

\begin{figure}[!t]
\centering
\IfFileExists{figures/mimic_v2_cfd_ranking.png}{\includegraphics[width=\linewidth]{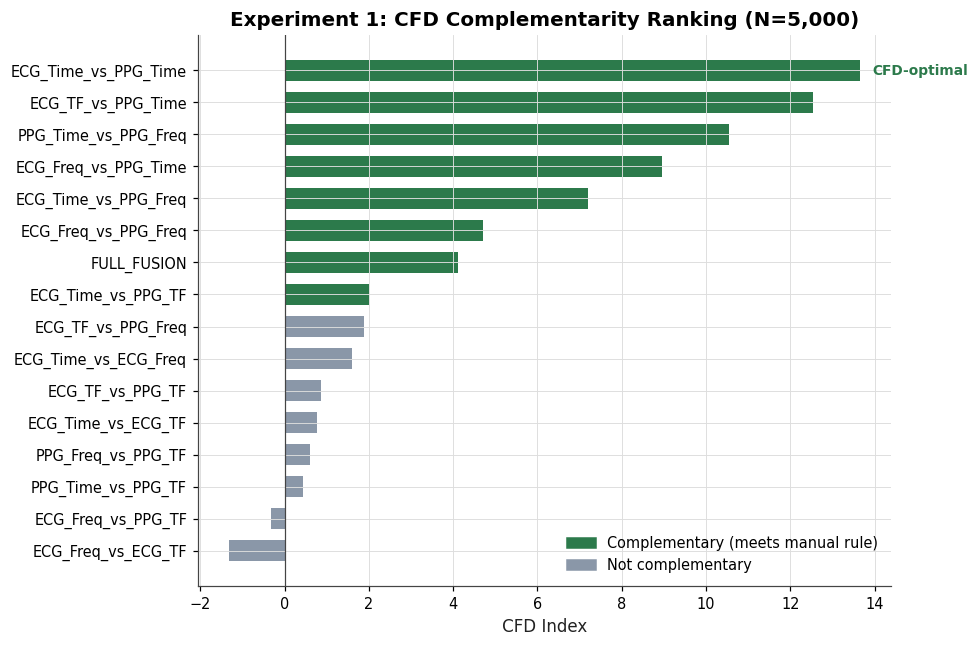}}{\figplaceholder{Upload MIMIC n=5,000 CFD ranking here as figures/mimic_v2_cfd_ranking.png}}
\caption{Experiment 7A: MIMIC n=5,000 CFD complementarity ranking across ECG and PPG time, frequency, and time-frequency domains.}
\label{fig:mimic_v2_cfd_ranking}
\end{figure}

Figure~\ref{fig:mimic_v2_cfd_ranking} complements the bootstrap results in Table~\ref{tab:mimic_stability_assignment}: ECG-Time+PPG-Time has the largest point estimate, but the broader conclusion is the separation between several positive cross-modal pairs and weak within-modality or time-frequency-heavy pairs. The ranking should therefore guide candidate selection and then be confirmed with resampling and held-out performance.

The MIMIC trust hierarchy refined the respiratory result. RESP was not used for CFD ranking or PECS training; it was reserved for \(C(t)=0\) ECG-PPG disagreement cases. On 626 disagreement cases, respiratory arbitration achieved TRA = 0.7588, exceeding the target of 0.70. However, applying the fixed trust hierarchy across the full test set reduced accuracy from fusion DCA = 0.9560 to 0.8704, with Level 1 agreement cases achieving 0.9840 accuracy, Level 2 RESP tiebreaker cases achieving 0.7561, and Level 3 uncertainty/fallback cases achieving 0.7273. This result supports respiration as a useful disagreement-case signal but not as a universal override; future versions should learn when RESP is trustworthy rather than applying it mechanically.

Morphology preservation also changed under the updated MIMIC run. Using the post-hoc PCA approximation of \(z_{\mathrm{stable}}\), CFD-optimal MPI averaged 0.4070, compared with 0.8500 for the redundant model and 0.7139 for the all-domain model. The resulting MPG values were negative (MPG = -0.4430 versus redundant, MPG = -0.3069 versus all domains), and the one-sided Wilcoxon tests did not support positive morphology gain after Bonferroni correction. As in the expanded BIDMC analysis, this indicates that morphology preservation should be interpreted together with DCA: the most reliable drift classifier can be more sensitive to clinically meaningful changes and therefore less cosine-stable under perturbation.

\begin{figure}[!t]
\centering
\IfFileExists{figures/mimic_v2_trust_flow.png}{\includegraphics[width=\linewidth]{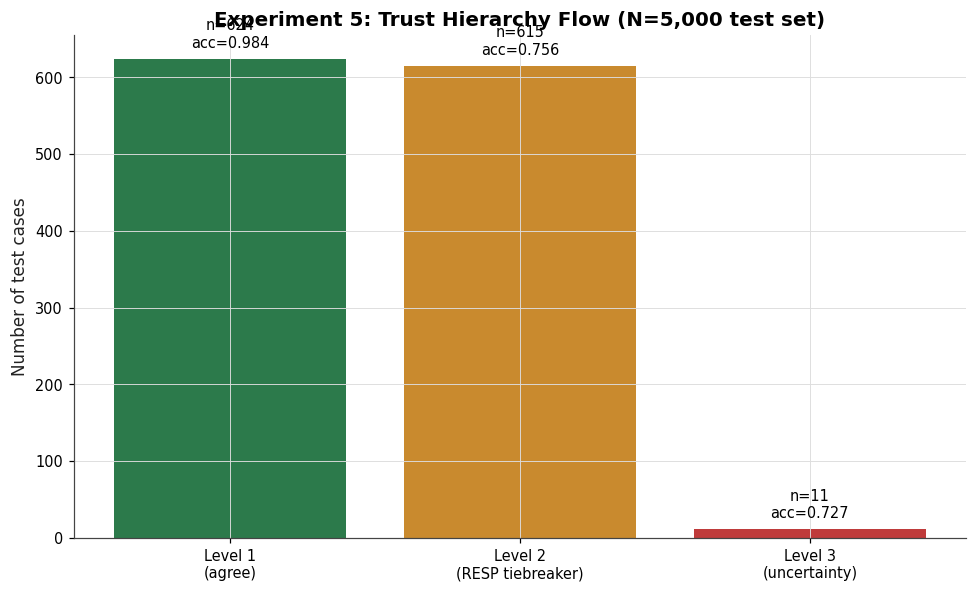}}{\figplaceholder{Upload MIMIC n=5,000 trust hierarchy flow here as figures/mimic_v2_trust_flow.png}}
\caption{Experiment 7B: MIMIC n=5,000 trust hierarchy flow showing Level 1 agreement, Level 2 RESP tiebreaking, and Level 3 uncertainty/fallback cases.}
\label{fig:mimic_v2_trust_flow}
\end{figure}

Figure~\ref{fig:mimic_v2_trust_flow} shows where the full-test reduction occurs: Level~1 agreement is highly reliable, whereas Level~2 respiratory arbitration and Level~3 fallback are weaker. This flow explains why a tiebreaker can exceed its local TRA target yet still reduce overall accuracy when applied mechanically.

\begin{figure}[!t]
\centering
\IfFileExists{figures/mimic_v2_mda_by_trust.png}{\includegraphics[width=\linewidth]{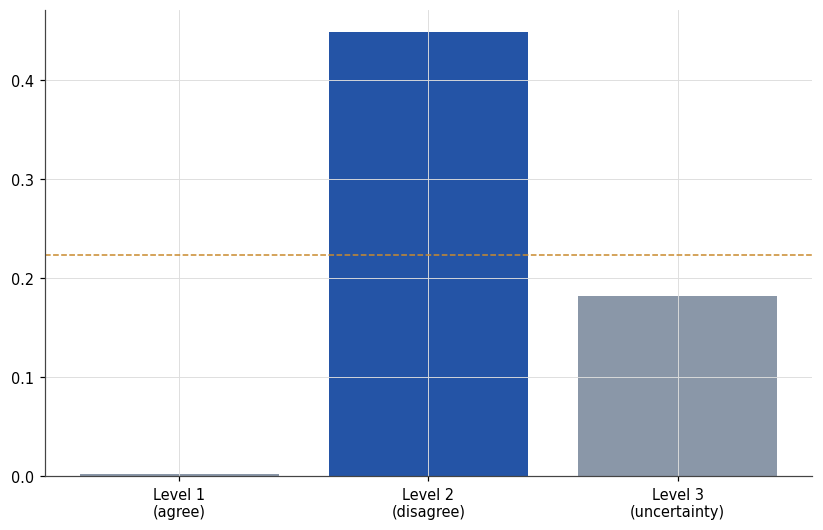}}{\figplaceholder{Upload MIMIC n=5,000 MDA by trust level here as figures/mimic_v2_mda_by_trust.png}}
\caption{Experiment 7C: MIMIC n=5,000 multimodal drift advantage by trust level.}
\label{fig:mimic_v2_mda_by_trust}
\end{figure}

Figure~\ref{fig:mimic_v2_mda_by_trust} further localizes the value of multimodality. The advantage is concentrated in conflict-sensitive trust levels rather than uniform across the decision path, reinforcing the proposed use of PPG and respiration as conditional evidence rather than mandatory inputs for every case.

\section{Discussion}
The results support the central claim that domain complementarity matters for drift-aware cardiovascular AI, but the comparative analysis also shows that complementarity is dataset-, scale-, and preprocessing-sensitive. In full-scale PTB-XL, Frequency+TimeFrequency was the only complementary pair and full-domain fusion was not complementary. In the 5,035-segment BIDMC analysis, ECG-Frequency+PPG-Time-Frequency was the recommended pair, whereas ECG-Time+PPG-Time ranked first in the MIMIC n=5,000 analysis. These differences show that CFD should be treated as a data-dependent discovery step rather than a fixed assumption. The stable finding across datasets is that selected domain pairs outperform weak or redundant pairs and that full-domain fusion is not automatically the safest or most interpretable solution. The smaller BIDMC feasibility result is retained in the supplementary material.

The MIMIC stability analysis strengthens this interpretation by showing that complementarity is a pairwise relationship, not a permanent property of a single signal domain. The same domain can be strong, weak, or unstable depending on its partner. ECG-Time and PPG-Time often act as bridging domains, but even they fail with some partners from the same modality or with time-frequency representations. This explains why heatmaps alone are incomplete: a heatmap shows magnitude, but the stability and non-transitivity analysis shows whether the relationship is reliable enough to guide model design.

The PECS drift results support H2 for the evaluated configurations. PECS substantially outperformed the reference implementations in both analyses, indicating that energy-geometric consistency is a useful inductive bias for physiologic signals. However, ADWIN and Page-Hinkley used library defaults, and the MMD and dropout comparators used fixed, prespecified configurations rather than a PECS-matched hyperparameter search. Generic statistical drift methods also target distribution change rather than the study's benign/non-benign physiologic label definition. The reported percentage-point gaps should therefore be interpreted as comparisons with these operating points, not as a universal ranking against optimally tuned generic detectors.

The methodological novelty lies in the integration and evaluation rather than in claiming CFD theory or the energy-geometric premise as new. Prior work introduced complementary feature-domain selection and motivated energy-constrained physiologic stability \cite{oladunni2025rethinking,oladunni2026energy,oladunni2025physiologic}. This study operationalizes those components as a cross-modal drift-monitoring architecture: CFD selects partner-dependent ECG--PPG domains, PECS makes modality-specific stability decisions, respiration is reserved for disagreement routing, and the combined policy is evaluated across PTB-XL, BIDMC, and MIMIC with perturbation stability, non-transitivity, morphology, and trust-level analyses. The new contribution is thus the multimodal decision framework and its cross-dataset evidence, not a reintroduction of the underlying CFD or energy formulations.

The perturbation-label design also limits the strength of the drift-classification claim. Although labels were assigned from perturbation identities before PECS features were calculated, the benign/non-benign taxonomy was designed using physiologic assumptions related to those encoded by PECS. The experiments therefore test whether PECS recovers a prespecified mechanistic distinction; they do not establish that it detects independently observed clinical deterioration. A stronger test would use blinded clinician adjudication, naturally occurring longitudinal state changes, or perturbation labels defined from outcomes unavailable to the PECS rule.

The morphology results qualify the reliability findings. BIDMC produced negative MPG under ST elevation and QRS widening, and MIMIC produced the same cautionary pattern: CFD-optimal fusion achieved strong drift reliability but lower post-hoc MPI than redundant and all-domain alternatives. This does not necessarily mean the CFD-optimal encoder is worse. The analyses indicate that CFD-optimal encoders can be more sensitive to clinical signal changes, producing higher DCA while preserving less cosine similarity under label-changing perturbations. Therefore, MPI should be interpreted together with DCA rather than as an isolated marker of clinical quality.

The trust hierarchy results are scenario-dependent. In BIDMC, TRA was 0.5133 but respiratory routing improved disagreement-case DCA by 16.23 percentage points. In MIMIC, TRA reached 0.7588 on 626 ECG-PPG disagreement cases, yet the fixed trust hierarchy reduced full-test accuracy from 0.9560 to 0.8704. This tension is informative: respiration can be valuable when ECG and PPG disagree, but it should be treated as a learned or confidence-weighted trust signal rather than as a mechanical override.

The expanded MDA experiment strengthens the case for conditional multimodality. Multimodal ECG+PPG drift detection improved substantially over ECG-only and PPG-only baselines, but the benefit concentrated in Level 2 disagreement cases. In Level 1 agreement cases, multimodal routing added little and slightly underperformed the best unimodal signal. This pattern supports the trust hierarchy: multimodal fusion is most useful when the modalities disagree, while agreement cases can often be handled more simply.

\section{Limitations}
This study has several limitations. First, the PTB-XL, BIDMC, and MIMIC experiments use different datasets, label constructions, and perturbation protocols, so cross-dataset comparisons should be interpreted cautiously. Second, benign/non-benign labels were assigned from synthetic perturbation families rather than independent clinical adjudication. Although labels were fixed before PECS energy and latent-drift features were computed, the taxonomy and decision rule share physiologic design assumptions, creating potential soft circularity. Third, the smaller and larger BIDMC analyses differ in sample size and preprocessing; the smaller feasibility analysis is therefore reported only in the supplementary material, and future validation should standardize the full pipeline. Fourth, the MIMIC trust hierarchy was evaluated on a 1,250-sample test partition after deriving 5,000 synchronized segments, so larger held-out validation is still needed. Fifth, some p-values are reported as zero by numerical routines and should be interpreted as \(p<10^{-6}\). Sixth, respiratory arbitration was evaluated with rule-based respiratory features rather than clinically adjudicated respiratory diagnoses. Finally, the current results support a controlled mechanistic research framework for drift-aware cardiovascular AI, not a deployable clinical device.

\section{Conclusion}
This paper presented a CFD-guided PECS framework for deciding when a cardiovascular AI model should hold, update, or flag its prediction under physiologic drift. Across PTB-XL, BIDMC, and MIMIC analyses, pyCFD identified targeted complementary domain pairs and repeatedly showed that full-domain fusion is not automatically complementary. PECS improved drift classification over the evaluated baselines, CFD-guided features improved reliability over redundant non-CFD features, and multimodal ECG+PPG drift detection provided the largest benefit during ECG-PPG disagreement. The MIMIC stability analysis further showed that complementarity can be non-transitive and partner-dependent, reinforcing the need for explicit domain selection rather than assuming a domain is always useful. The comparison also revealed important limits: the highest-ranked CFD pair changed across datasets, morphology preservation can trade off against clinical sensitivity, and respiration is useful only when treated as a selective or learned trust signal. Overall, the findings support CFD-guided PECS as a candidate monitoring framework while motivating standardized large-scale ECG-PPG-RESP validation and clinically adjudicated drift labels.

\balance
\bibliographystyle{IEEEtran}
\bibliography{references}

\end{document}